\documentclass[aps, reprint,pra,superscriptaddress,nofootinbib,longbibliography]{revtex4-1}
\usepackage{dcolumn}
\usepackage[utf8]{inputenc}
\usepackage{amsmath}
\usepackage{amsfonts}
\usepackage{amssymb}
\usepackage{graphicx}
\usepackage{hyperref}
\usepackage{xcolor}

\newcommand{\mb}{\mathbf}

\begin{document}

\title{Topological schemes in spacetime for the electrodynamic Aharonov-Bohm effect}

\author{P. L. Saldanha}\email{saldanha@fisica.ufmg.br}
\affiliation{Departamento de F\'isica, Universidade Federal de Minas Gerais, Belo Horizonte, MG 31270-901, Brazil}

\author{H. Batelaan}
\affiliation{Department of Physics and Astronomy, University of Nebraska-Lincoln, 208 Jorgensen Hall, Lincoln, Nebraska 68588-0299, USA}

\date{\today}

\begin{abstract}
We consider different schemes for the electrodynamic Aharonov-Bohm (AB) effect introduced in Ref. [Phys. Rev. A $\mathbf{108}$, 062218 (2023)], exploring the phenomenon to enhance the understanding of its topological nature in spacetime. In the treated examples, the electric current in a solenoid varies in time, changing its internal magnetic field and producing an external electric field, while a quantum charged particle is in a superposition state inside two Faraday cages in an interferometer. The Faraday cages cancel the electric field at their interiors, such that the particle is always subjected to null electromagnetic fields. We discuss how the AB phase difference depends on the topology of the electric and magnetic fields in spacetime in the different treated situations. In particular, we discuss interesting results when a conducting wire connects the two Faraday cages, with the AB phase depending on the wire position. We also show an amplification of the AB phase when the wire makes several turns around the solenoid, which could enable an experimental verification of the effect.
\end{abstract}


\maketitle

\section{Introduction} The Aharonov-Bohm (AB) effect \cite{ehrenberg49,aharonov59} is a striking quantum phenomenon that contradicts the idea that electric charges are affected only by the local electromagnetic fields. In the magnetic AB effect \cite{ehrenberg49,aharonov59}, the interference pattern of quantum charged particles in an interferometer, whose possible paths enclose a long solenoid, depends on the solenoid's magnetic flux even if the particles only propagate in regions with negligible electromagnetic fields. In the electric AB effect \cite{aharonov59}, the interference pattern of quantum charged particles depends on the electric potentials applied to the different interferometer paths during the particles' passage even if the particles are always subjected to negligible electromagnetic fields. Several experiments performed with different systems have confirmed the AB effect \cite{chambers60,matteucci85,webb85,tonomura86,oudenaarden98,bachtold99,ji03,peng10,becker19,nakamura19,deprez21,ronen21}.

The magnetic AB effect is frequently described as a topological effect \cite{peshkin95,cohen19,ballentine,bransden,gottfried}, with the AB phase depending on the magnetic flux enclosed by the quantum particle's possible trajectories in the interferometer. A description of the electric AB effect was recently given, with the AB phase depending on the electric flux in a spacetime surface whose boundaries are the quantum particle's possible world lines in the interferometer \cite{saldanha23}. A mathematically equivalent view is that the AB phase depends on the topologies of the electromagnetic field configuration and possible particles' trajectories in spacetime. This latter description permitted the prediction of a novel electrodynamic AB effect, where a nonzero AB phase difference appears in a situation where the magnetic flux in a solenoid varies while the quantum particle is in a superposition state inside two Faraday cages in an interferometer, even if the particle paths enclose no magnetic flux and are subjected to a negligible scalar potential difference \cite{saldanha23}. 

Here we present and explore the electrodynamic AB effect introduced in Ref. \cite{saldanha23} in novel situations, providing a deeper understanding of the AB effect, regarding its topological nature. All the scenarios considered are type-I AB-effects. A non-zero phase shift is accumulated in an interferometer, while the interfering particle never interacts locally with an electric or magnetic field.   All the scenarios discussed are topological in space-time and are explicit examples of the electrodynamic AB-effect. This can be distinguished from the original magnetic AB-effect, where fields are time-independent, the effect is topological in space and does not need the adjective ``electrodynamic''. Our exposition moves in methodical steps from the example from our earlier work in Ref. \cite{saldanha23}, where a magnetic flux tube is not enclosed and still gives a phase shift, to an example that is recognizable as an original AB-effect where an electric flux is enclosed and that could be performed in an experiment. First the solenoid is moved back into the interferometer, second, a conducting wire is added in two ways to highlight topology, third, the wire is extended and the solenoid is moved back outside the interferometer.  In the final scenario, the conducting wire makes $N$ turns around the solenoid. For a magnetic field that is varied, the AB phase is amplified by a factor of $N$. This could facilitate an experimental verification of the electric AB-effect for free electrons, the demonstration of which has remained elusive. 

\section{First topological scheme}
Consider the situation depicted in Fig. 1(a), with a Mach-Zehnder interferometer for a non-relativistic quantum charged particle. The initial magnetic flux in the infinite solenoid is $\Phi_i$. While the quantum particle is in a superposition state inside both Faraday cages, which are symmetrically positioned in relation to the solenoid as shown in the figure, its electric current varies, with the final magnetic flux being $\Phi_f$. This change of the electric current produces an electric field in the interferometer, but the Faraday cages cancel the fields at their interior, such that the quantum particle is subjected to null electromagnetic fields. The quantum particle then continues its evolution through the interferometer, always in regions with null electromagnetic fields. 

\begin{figure}
  \centering
    \includegraphics[width=8.5cm]{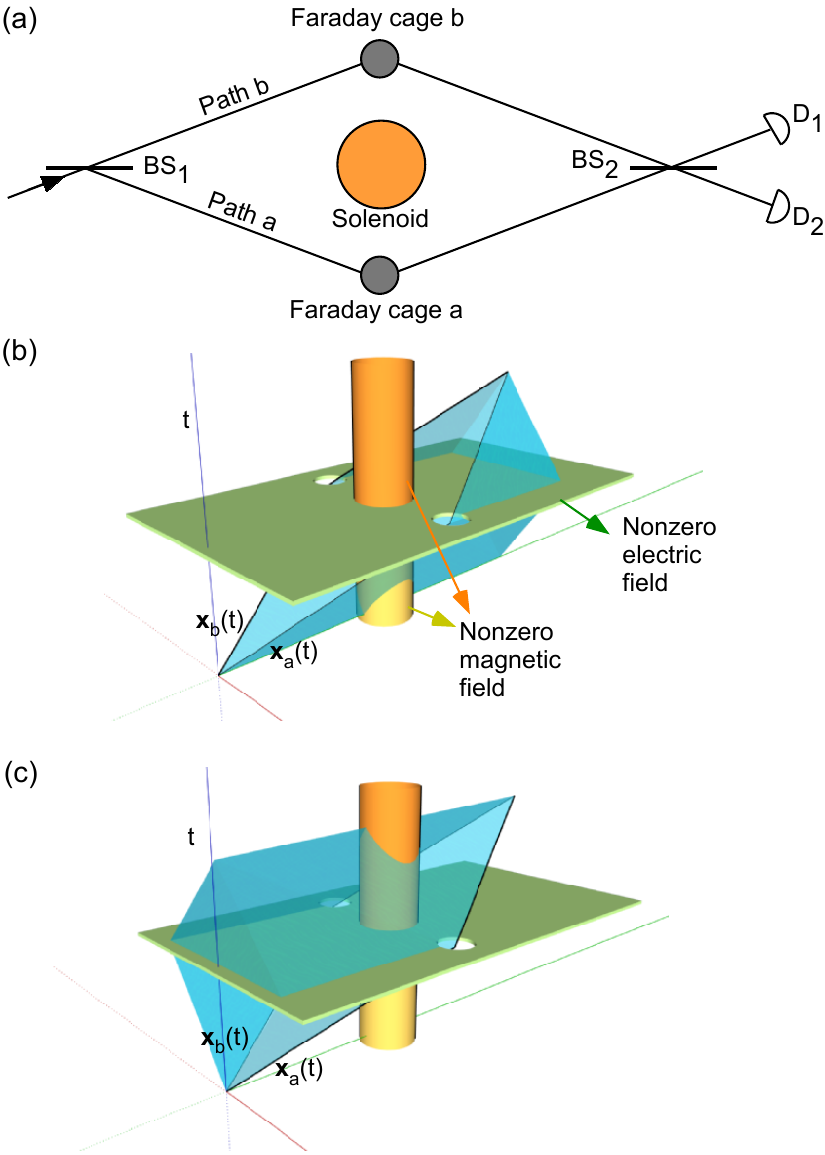}
  \caption{(a) Scheme of an electrodynamic AB interferometer for a quantum charged particle. BS$_1$ and BS$_2$ are beam splitters, D$_1$ and D$_2$ are detectors. The particle is sent to the interferometer as indicated by the arrow. While the quantum particle is in a superposition state inside both Faraday cages, the solenoid current is changed. (b) The possible spacetime trajectories $\mb{x}_a(t)$ and $\mb{x}_b(t)$ of the quantum particle in the interferometer are represented in black. The time coordinate is in the vertical direction and the trajectories are in the $xy$ plane. The spacetime regions with nonzero magnetic field are represented in orange and yellow, with these two colors representing two different values for the field. The spacetime region with nonzero electric field is represented in green. The spacetime surface formed by one possible deformation of the spacetime trajectory $\mb{x}_a(t)$ into the spacetime trajectory $\mb{x}_b(t)$ is represented in blue. (c) The same as in (b), but with a different spacetime surface in blue formed by a different deformation of the spacetime trajectory $\mb{x}_a(t)$ into the spacetime trajectory $\mb{x}_b(t)$.}
\end{figure}

In Ref. \cite{saldanha23}, following Ref. \cite{singleton13}, it was discussed how the AB phase difference between the interferometer paths in any AB scheme can be written as
\begin{equation}\label{AB_gen}
	\phi_{AB}=\frac{q}{\hbar}\left[\int_\mathcal{S}d\mb{a}(t)\cdot\mb{B}(\mb{r},t)-\int  \int_\mathcal{S} dt\,  d\mb{r}(t)\cdot\mb{E}(\mb{r},t)\right],
\end{equation}
where $q$ is the particle charge, $\mb{E}$ is the electric field, $\mb{B}$ is the magnetic field, and $\mathcal{S}$ is a spacetime surface bounded by the world lines of the two interfering paths. This surface can be constructed from the deformation of one of the particle possible trajectories in the interferometer in spacetime into the other, in the following way. Consider that at each time instant $t$ the quantum particle is in a superposition state with wavepackets centered at two positions $\mb{x}_a(t)$ (in path $a$) and $\mb{x}_b(t)$ (in path $b$). Draw an arbitrary curve that goes from $\mb{x}_a(t)$ to $\mb{x}_b(t)$ for each time instant $t$, through infinitesimal displacements $d\mb{r}(t)$. The complete spacetime surface is constructed by connecting these curves at different times. More details can be found in Ref. \cite{saldanha23}. One example of such spacetime surface is the blue surface in Fig. 1(b). Another equally valid example of such spacetime surface, with the same boundary, is the blue surface in Fig. 1(c). The spacetime surfaces used in each of the examples that we present were chosen to simplify the calculations.

In Fig. 1(b), the two possible particle trajectories in spacetime are represented in black. The initial magnetic field is represented in yellow, while the final one is in orange. The spacetime region with a nonzero electric field is represented in green. It can be seen that the region with null electromagnetic fields is not simply connected in spacetime. It is not possible to deform one of the possible particle trajectories in spacetime into the other, keeping the initial and final points fixed, without crossing a region with nonzero electromagnetic fields. The AB phase difference of Eq. (\ref{AB_gen}) depends on the topologies of the electromagnetic fields and possible particle trajectories in spacetime, not on specific geometric configurations \cite{saldanha23}. If the particle's trajectories are varied, but with one path crossing one of the ``holes'' in the figure and the other path crossing the other ``hole'', the AB phase difference does not change. A peculiarity of the topology of the electrodynamic (and also of the electric) AB effect is that the winding number is always 1. For having a winding number greater than one, the particle world line in one path would need to cross one of the ``holes'' in Fig. 1(b), to come back in time crossing the other ``hole'', and to cross the first ``hole'' again. Since it is not possible for the quantum particle to travel backwards in time, the winding number is always 1 in this case. Note that we will consider schemes for which the interferometer lengths are small and the timescales are large, such that the fields propagation times within the interferometer can be disregarded.

Now let us compute the AB phase difference between the paths in the scheme of Fig. 1(a) using Eq. (\ref{AB_gen}) with the spacetime surface represented in Fig. 1(b). Note that the initial magnetic flux crosses the spacetime surface of Fig. 1(b), such that the contribution of the first term on the right side of Eq. (\ref{AB_gen}) to the AB phase difference is $q\Phi_i/\hbar$. In the Lorenz gauge, a long cylindrical solenoid with a magnetic flux $\Phi$ produces a vector potential $\mb{A}=\Phi/(2\pi\rho)\boldsymbol{\hat{\phi}}$ outside it, in cylindrical coordinates $(\rho,\phi,z)$. While the solenoid current varies, it generates an electric field $\mb{E}_A=-\partial\mb{A}/\partial t$ in the $\boldsymbol{\hat{\phi}}$ direction. Charge densities are induced in the Faraday cages, creating an electric field $\mb{E}_V$ that cancels the total electric field at their interior. We consider that the center of the Faraday cages are in positions $\mb{R}_j$, with $j=\{a,b\}$. The total electric field can be written as $\mb{E}=\mb{E}_A+\mb{E}_V$, which enters in the the second term on the right side of Eq. (\ref{AB_gen}) contributing to the AB phase, since we have a nonzero electric flux in the spacetime surface represented in blue in Fig. 1(b). This electric flux occurs in the superposition of this spacetime surface with the green region, where the electric field is nonzero.

Due to the system symmetry, the induced charges are symmetric with a rotation of 180 degrees around the solenoid axis, such that the electric field $\mb{E}_V$ produced by these induced charges is also symmetric with this rotation.  
Any closed loop integral at constant t, $\oint d\mb{r}\cdot\mb{E}_V(\mb{r})$, for the field $\mb{E}_V$, due to the induced charge distribution, equals zero. Combined with rotational symmetry of $\mb{E}_V$, the line integral $\int_{\mb{R}_a}^{\mb{R}_b}\mb{E}_V\cdot d\mb{r}$ over the line defined by the intersection of the green and blue surface at any time during which the electron is in the Faraday cage in Fig. 1(b) equals zero. So, $\mb{E}_V$ does not contribute to the AB phase of Eq. (\ref{AB_gen}) in this case.   

But we do have a contribution of the field $\mb{E}_A=-\partial\mb{A}/\partial t$ to the electric flux in the spacetime surface in blue in Fig. 1(b), contributing to the AB phase of Eq. (\ref{AB_gen}).  The nonzero contribution of the time integral in Eq. (\ref{AB_gen}) goes from a time $t_0$, when the current in the solenoid starts to vary, producing an electric field, to a time $t_1$, when it becomes stationary again and the electric field disappears. The spatial integral goes from a position $\mb{R}_a$ in cage $a$ to a position $\mb{R}_b$ in cage $b$ through the blue surface of Fig. 1(b) at each time instant $t$. So, the contribution of the second term on the right side of Eq. (\ref{AB_gen}) to the AB phase difference is
\begin{equation}\label{aux}
	\frac{q}{\hbar}\int_{t_0}^{t_1} dt\int_{\mb{R}_a}^{\mb{R}_b}  d\mb{r}\cdot\frac{\partial\mb{A}}{\partial t}=\frac{q}{\hbar}\left[ \int_{\mb{R}_a}^{\mb{R}_b} d\mb{r}\cdot\mb{A}\right]_{t_0}^{t_1}.
\end{equation}
At time $t_0$, we have $\mb{A}_i=\Phi_i/(2\pi\rho)\boldsymbol{\hat{\phi}}$. Dividing the path integral above from $\mb{R}_a$ to $\mb{R}_b$ into infinitesimal dislocations in the directions $\boldsymbol{\hat{\rho}}$ and $\boldsymbol{\hat{\phi}}$ along the blue surface of Fig. 1(b) at time $t_0$, such that only the dislocations in the direction $\boldsymbol{\hat{\phi}}$ contribute for the integral, we have $\int_{\mb{R}_a}^{\mb{R}_b} d\mb{r}\cdot\mb{A}_i=\Phi_i/2$, considering a negligible size for the Faraday cages. The path integral at time $t_f$ can be computed in an analogous way, such that Eq. (\ref{aux}) becomes $q(\Phi_f-\Phi_i)/(2\hbar)$. If the Faraday cages are not so small, this result would be multiplied by  a numerical factor of the order of $(\pi-\Delta\theta)/\pi$, where $\Delta\theta$ is the angle formed by the lines that go from the solenoid center to each of the extremities of one of the cages in Fig. 1(a). 

So, since the first term on the right side of Eq. (\ref{AB_gen}) contributes with $q\Phi_i/\hbar$ and the second term with $q(\Phi_f-\Phi_i)/(2\hbar)$ for the AB phase, we conclude that the total AB phase difference between the paths in the scheme of Fig. 1(a) is $q(\Phi_i+\Phi_f)/(2\hbar)$. This result makes sense, since the total AB phase difference depends on the average magnetic flux enclosed by the paths during the particle propagation through the interferometer. During half of the path the particle is subjected to the vector potential $\mb{A}_i$, and half of the path to the potential $\mb{A}_f$.

If we consider the spacetime surface of Fig. 1(c), the contribution of the first term on the right side of Eq. (\ref{AB_gen}) is $q\Phi_f/\hbar$ and the contribution of the second term is $q(\Phi_i-\Phi_f)/(2\hbar)$, resulting in the same total AB phase difference. The sign of the latter contribution has changed as the direction of the path integration starting at $\mb{R}_{a}$ and ending at $\mb{R}_{b}$ has changed in Eq. (\ref{aux}), being clockwise instead of counterclockwise. Note that in the scheme of Fig. 1(a) both terms of Eq. (\ref{AB_gen}) contribute for the total AB phase difference, no matter what spacetime surface is constructed, as long as $\Phi_i$ and $\Phi_f$ have different values, both different from zero. This wasn't true in the examples treated in Ref. \cite{saldanha23}, where the particle paths did not enclose the solenoid. In the examples treated there, for specific spacetime surfaces one of the two terms on the right side of Eq. (\ref{AB_gen}) could have a null contribution for the AB phase. This is an important novelty of the AB scheme of Fig. 1(a) treated here. To our knowledge, it is the first AB scheme where both the electric and magnetic field configurations necessarily contribute for the total AB phase difference between the interferometer paths, being a truly \textit{electromagnetic} AB effect.  

\section{Second topological scheme}
Now let us consider the scheme of Fig. 2(a), similar to the one of Fig. 1(a). Again, the magnetic flux in the infinite solenoid varies from an initial value $\Phi_i$ to a final value $\Phi_f$ while the quantum particle is in a superposition state inside the Faraday cages. But now we have a metallic wire connecting the cages, as shown in the figure. Fig. 2(b) illustrates the two interferometric particle trajectories and the electromagnetic field configuration in spacetime. The total AB phase difference between the paths can be computed using Eq. (\ref{AB_gen}) with the spacetime surface indicated in blue in Fig. 2(b). The contribution of the second term on the right side of Eq. (\ref{AB_gen}) is zero for this spacetime surface, since there is no superposition of the blue surface with the green solid in this case (or, in other words, $\mb{E}$ is zero over the surface of integration). The contribution of the first term is $q\Phi_i/\hbar$, which becomes the total AB phase difference. Interestingly, the AB phase difference does not depend on the final solenoid magnetic flux. The opposite happens in the scheme of Fig. 2(c), with the metallic wire turned to the other side of the solenoid. In an analogous way to Fig. 2(b), we can consider a spacetime surface with no superposition with the green solid, such that the second term on the right side of Eq. (\ref{AB_gen}) does not contribute for the AB phase difference. However, in this case the first term results in $q\Phi_f/\hbar$. So, in this situation the AB phase difference is independent from the initial solenoid magnetic flux. This is an interesting subtlety of the effect: The AB phase difference depends on the position of the metallic wire that connects the Faraday cages.

\begin{figure}
  \centering
    \includegraphics[width=8.5cm]{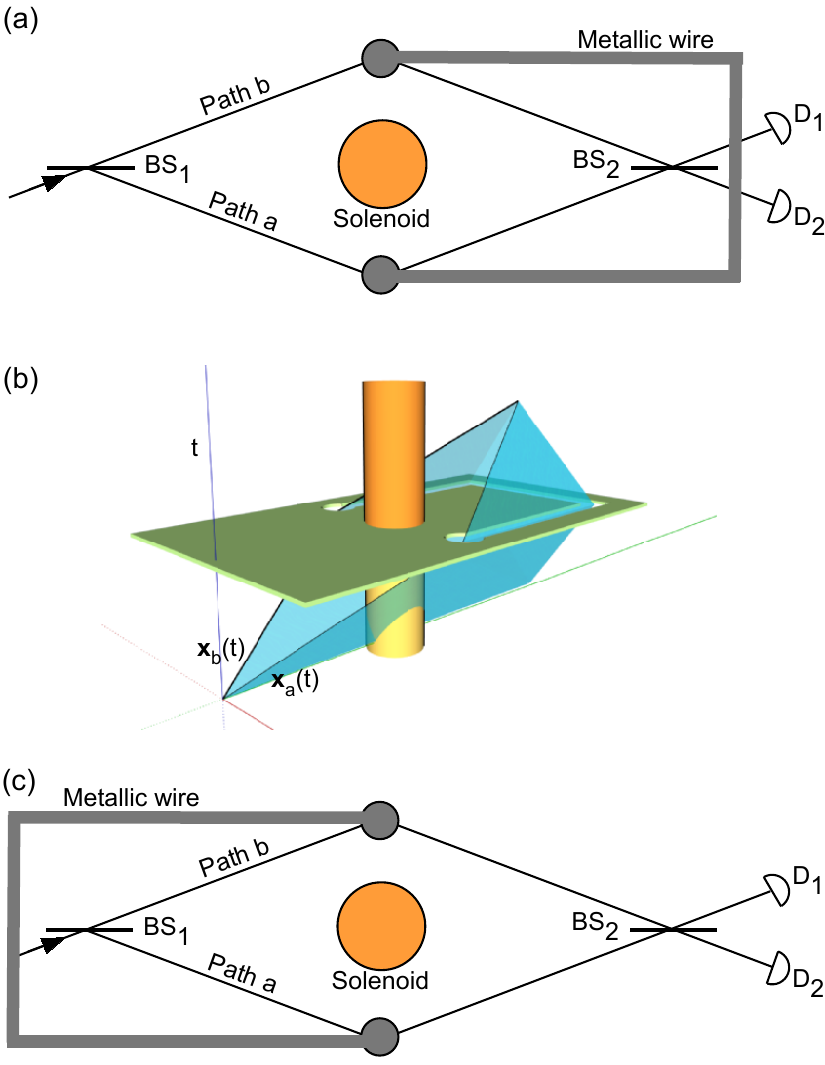}
  \caption{(a) An interferometer similar to the one of Fig. 1(a), but now with a metallic wire connecting the two Faraday cages. Again, the solenoid current varies while the quantum particle is in a superposition state inside the Faraday cages. (b) Representation of the two interferometric particle trajectories and electromagnetic fields in spacetime, similarly to Fig. 1(b). Note that the spacetime surface (blue) formed by the deformation of one spacetime trajectory into the other has no superposition with the region with nonzero electric field (green) in this case. (c) The metallic wire of panel (a) is now turned to the other side of the solenoid.}
\end{figure}

The enhance the understanding of the above example, it can be compared to its description in a more conventional approach \cite{aharonov59}. Consider the continuous accumulation of the AB phase difference acquired by the quantum particle during its propagation along its path through the interferometer, a view supported by works that use a quantum electromagnetic field to describe the interactions in the AB effect \cite{santos99,marletto20,saldanha21a,saldanha21b,saldanha24}. 
The AB phase results from the term $H_I = qV-(q/m)\mb{p}\cdot\mb{A}$ in the system Hamiltonian, where $q$, $m$, and $\mb{p}$ are the particle charge, mass, and momentum respectively, and $V$ and $\mb{A}$ are  the scalar and vector potentials. The AB phase accumulated in each path can be written as
\begin{equation}\label{AB_i}
	\phi_{i}=-\int_{t_0}^t \frac{H_I}{\hbar} dt'=-\frac{q}{\hbar}\int_{t_0}^t Vdt' +\frac{q}{\hbar}\int_{\mb{x}_0}^{\mb{x}_i}\mb{A}\cdot d\mb{x},
\end{equation}
where $\mb{p}/m$ was substituted by the average wave packet velocity $d\mb{x}_i/dt$, we have $\mb{x}_a(t_0)=\mb{x}_b(t_0)=\mb{x}_0$ before the wave function splitting in the interferometer, and the spatial integral is performed through the particle path. 

In the scheme of Fig. 2(a), while the particle propagates from BS$_1$ to the Faraday cages, the contribution for the AB phase difference can be computed by the second term on the right side of Eq. (\ref{AB_i}). In the Lorenz gauge, we have $\mb{A}=\Phi/(2\pi\rho)\boldsymbol{\hat{\phi}}$ in cylindrical coordinates. Dividing the path integral into infinitesimal dislocations in the directions $\boldsymbol{\hat{\rho}}$ and $\boldsymbol{\hat{\phi}}$, we conclude that the contribution of the second term on the right side of Eq. (\ref{AB_i}) for the AB phase difference is $q\Phi_i/(2\hbar)$ for very small cages. While the magnetic flux in the solenoid varies, an electric field $\mb{E}_A=-\partial\mb{A}/\partial t$ is created around it. Electric charges are then induced in the Faraday cages and in the metallic wire creating an electric field $\mb{E}_V=-\mb{E}_A$ inside them, canceling the total field. This generates a potential difference $V_a-V_b=-\int_{\mb{R}_b}^{\mb{R}_a} \mb{E}_V\cdot d\mb{r}=\int_{\mb{R}_b}^{\mb{R}_a}  d\mb{r}\cdot\partial\mb{A}/\partial t$ between the Faraday cages. Following the same steps that were used after Eq. (\ref{aux}), we see that the first term on the right side of Eq. (\ref{AB_i}) results in a phase difference $q(\Phi_i-\Phi_f)/(2\hbar)$ during the variation of the solenoid magnetic flux. Finally, while the quantum particle propagates from the Faraday cages to BS$_2$, we have a contribution $q\Phi_f/(2\hbar)$ for the AB phase difference, that comes from the second term on the right side of Eq. (\ref{AB_i}). The three terms together results in a total AB phase difference $q\Phi_i/\hbar$, as previously deduced with the use of Eq. (\ref{AB_gen}). Note that in the scheme of Fig. 2(c) the potential difference between the cages while the solenoid magnetic flux varies changes its sign, since the direction of $\mb{E}_A$ (and consequently of $\mb{E}_V$) is inverted in opposite sides of the solenoid. This results in a total AB phase difference $q\Phi_f/\hbar$ in the scheme of Fig. 2(c), as previously computed.

We thus see that the AB phase difference between the paths in the schemes shown in Fig. 2 depends on the position of the metallic wire connecting the cages. This interesting behavior can be readily predicted with the spacetime diagrams as the one of Fig. 2(b). These examples show how these spacetime diagrams are useful for the prediction of novel interesting behaviors, besides evidencing the topological nature of the different kinds of AB effects.

\section{Third topological scheme}

Let us consider a last example, represented in Fig. 3, with the solenoid outside the interferometer, as in the situation considered in Ref. \cite{saldanha23}. But now there is a metallic wire connecting the Faraday cages, giving $N$ turns around the solenoid (with $N=2$ in this figure). Again, the solenoid magnetic flux varies while the quantum particle is in a superposition state inside both Faraday cages. Electric charge densities are induced in the Faraday cages and in the metallic wire canceling the total electric field inside them. So, inside the wire the electric field generated by the induced charges is given by $\mb{E}_V=-\mb{E}_A=\partial \mb{A}/\partial t$. The potential difference between the Faraday cages can be written as $V_a-V_b=-\int_{\mb{R}_b}^{\mb{R}_a} \mb{E}_V\cdot d\mb{r}$ in a path inside the metallic wire. Using this result in the first term of the right side of Eq. (\ref{AB_i}), we obtain a total AB phase difference $\pm Nq(\Phi_i-\Phi_f)/\hbar$, since the second term results in a null contribution in this configuration, and where the sign is determined by the direction of the wire turns, consistent with the ``half'' number of turns of the example of Figure 2. We have an amplification effect on this electromagnetic AB phase difference given by the number of turns $N$ of the metallic wire around the solenoid. In addition to illustrating a topological scheme, the amplification may be of use for the observation of the dispersionless nature of the Aharonov-Bohm effect for free electrons, which has remained elusive \cite{batelaantonomura}. Since 3 spatial dimensions are necessary to properly represent this scheme, with the metallic wire giving $N$ turns around the solenoid, it is not possible to construct spacetime diagrams like the ones from Figs. 1(b), 1(c) and 2(b) in this case, since it would be necessary 3 spatial plus 1 time dimensions. 

\begin{figure}
  \centering
    \includegraphics[width=8.5cm]{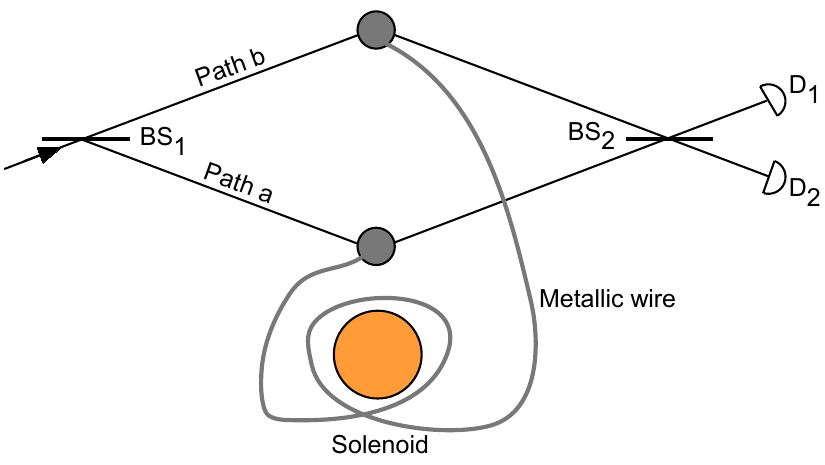}
  \caption{Scheme of an electrodynamic AB interferometer for a quantum charged particle similar to the one of Fig. 1(a), but with the solenoid outside the interferometer. A metallic wire connects the two Faraday cages with $N=2$ turns around the solenoid. Again, the solenoid current is changed while the quantum particle is in a superposition state inside both Faraday cages.}
\end{figure}

\section{Discussion and Conclusion}

The traditional magnetic AB-effect describes a charged particle interferometer that encloses a magnetic flux in space and causes an AB-phase shift. For the archetypical case when a magnetic flux tube is enclosed, the phase shift does not depend on the location of the magnetic flux tube within the interferometer. When the flux tube is not enclosed the phase shift is zero independent of location. There is thus a direct correlation between the topology of the physical system and the measured AB-effect. The topology is defined in the sense that the electron interferometer arms cannot be continuously deformed in space from the enclosing geometry to the non-enclosing geometry without passing through the flux tube. Or, equivalently, one cannot move the solenoid so that the phase shift is removed, while keeping the topology the same.  
In  Ref. \cite{saldanha23}, it was shown that a magnetic flux tube that is not enclosed but switches in time, can still lead to an AB-effect, an example of a type-I AB effect, that is topological in space-time. To expand on this, the effect can be understood as topological when considering the electron interferometer arms in space-time. In space-time, the arms cannot be continuously deformed or transformed without passing through a flux (in this case the electric flux between the Faraday cages). Now, let’s consider similar interferometer-arm transformations for the scenarios described in our manuscript. To move from the second scenario in our manuscript, depicted in Figs. 2(a) and 2(b), to the scenario considered in  Ref. \cite{saldanha23} with the solenoid outside the interferometer (with an additional wire connecting the Faraday cages), first choose the initial magnetic flux [yellow region in Fig. 2(b)] to be zero. Now move the solenoid away from the conducting wire, in the direction of BS$_1$, through the interferometer arms to outside the interferometer. Given that a null flux  passes through the possible particle trajectories $\mb{x}_a(t)$ and $\mb{x}_b(t)$ in Fig. 2(b), this does not constitute a change and the topology remains the same. In other words, the topology of the example of Ref. \cite{saldanha23} modified with a conducting wire and with $\Phi_i = 0$, is topologically equivalent to the second example in our manuscript, represented in Figs. 2(a) and 2(b), with $\Phi_i = 0$. Given that the effect is purely topological and these two examples are topologically identical, the phase shift should be the same. The phase shift in the first case is proportional to the electric flux on the spacetime surface of Fig. 3(b) from Ref. \cite{saldanha23}. Adding the conducting wire sets this electric flux to zero, resulting in a null phase, while for the second case it is $\frac{q}{\hbar}(\Phi_{i})$ . When $\Phi_i = 0$, these answers are the same, as they should be, illustrating topological equivalence in space-time. Alternatively, one can continuously deform the $t>0$ possible trajectories $\mb{x}_a(t)$ and $\mb{x}_b(t)$ in Fig. 2(b) to pass through the field-free regions in the conducting wire and then the $t<0$ possible trajectories $\mb{x}_a(t)$ and $\mb{x}_b(t)$ through the solenoid (when the flux is off) to show topological equivalence.  The same reasoning shows that the example depicted in Fig. 2(c) from our manuscript depends on $\Phi_{f}$ (the flux at $t>0$) and is topological distinct from the case of Fig. 2(a). 

To conclude, we have presented different physical schemes of the electrodynamic AB effect to illustrate what may be consider counter-intuitive behaviors associated with the AB effect and to reinforce its topological nature. The AB phase difference can always be computed by the general form of Eq. (\ref{AB_gen}), depending on the topologies of the electromagnetic fields and of the particles trajectories in spacetime.

\section{Acknowledgements}
PLS was supported by the Brazilian agency CNPq (Conselho Nacional de Desenvolvimento Cient\'ifico e Tecnol\'ogico). HB was supported by NSF award nr. 2207697.


\begin{thebibliography}{27}%
\makeatletter
\providecommand \@ifxundefined [1]{%
 \@ifx{#1\undefined}
}%
\providecommand \@ifnum [1]{%
 \ifnum #1\expandafter \@firstoftwo
 \else \expandafter \@secondoftwo
 \fi
}%
\providecommand \@ifx [1]{%
 \ifx #1\expandafter \@firstoftwo
 \else \expandafter \@secondoftwo
 \fi
}%
\providecommand \natexlab [1]{#1}%
\providecommand \enquote  [1]{``#1''}%
\providecommand \bibnamefont  [1]{#1}%
\providecommand \bibfnamefont [1]{#1}%
\providecommand \citenamefont [1]{#1}%
\providecommand \href@noop [0]{\@secondoftwo}%
\providecommand \href [0]{\begingroup \@sanitize@url \@href}%
\providecommand \@href[1]{\@@startlink{#1}\@@href}%
\providecommand \@@href[1]{\endgroup#1\@@endlink}%
\providecommand \@sanitize@url [0]{\catcode `\\12\catcode `\$12\catcode
  `\&12\catcode `\#12\catcode `\^12\catcode `\_12\catcode `\%12\relax}%
\providecommand \@@startlink[1]{}%
\providecommand \@@endlink[0]{}%
\providecommand \url  [0]{\begingroup\@sanitize@url \@url }%
\providecommand \@url [1]{\endgroup\@href {#1}{\urlprefix }}%
\providecommand \urlprefix  [0]{URL }%
\providecommand \Eprint [0]{\href }%
\providecommand \doibase [0]{http://dx.doi.org/}%
\providecommand \selectlanguage [0]{\@gobble}%
\providecommand \bibinfo  [0]{\@secondoftwo}%
\providecommand \bibfield  [0]{\@secondoftwo}%
\providecommand \translation [1]{[#1]}%
\providecommand \BibitemOpen [0]{}%
\providecommand \bibitemStop [0]{}%
\providecommand \bibitemNoStop [0]{.\EOS\space}%
\providecommand \EOS [0]{\spacefactor3000\relax}%
\providecommand \BibitemShut  [1]{\csname bibitem#1\endcsname}%
\let\auto@bib@innerbib\@empty
\bibitem [{\citenamefont {Ehrenberg}\ and\ \citenamefont
  {Siday}(1949)}]{ehrenberg49}%
  \BibitemOpen
  \bibfield  {author} {\bibinfo {author} {\bibfnamefont {W}~\bibnamefont
  {Ehrenberg}}\ and\ \bibinfo {author} {\bibfnamefont {R~E}\ \bibnamefont
  {Siday}},\ }\bibfield  {title} {\enquote {\bibinfo {title} {The refractive
  index in electron optics and the principles of dynamics},}\ }\href {\doibase
  10.1088/0370-1301/62/1/303} {\bibfield  {journal} {\bibinfo  {journal} {Proc.
  Phys. Soc. B}\ }\textbf {\bibinfo {volume} {62}},\ \bibinfo {pages} {8}
  (\bibinfo {year} {1949})}\BibitemShut {NoStop}%
\bibitem [{\citenamefont {Aharonov}\ and\ \citenamefont
  {Bohm}(1959)}]{aharonov59}%
  \BibitemOpen
  \bibfield  {author} {\bibinfo {author} {\bibfnamefont {Y.}~\bibnamefont
  {Aharonov}}\ and\ \bibinfo {author} {\bibfnamefont {D.}~\bibnamefont
  {Bohm}},\ }\bibfield  {title} {\enquote {\bibinfo {title} {Significance of
  electromagnetic potentials in the quantum theory},}\ }\href {\doibase
  10.1103/PhysRev.115.485} {\bibfield  {journal} {\bibinfo  {journal} {Phys.
  Rev.}\ }\textbf {\bibinfo {volume} {115}},\ \bibinfo {pages} {485} (\bibinfo
  {year} {1959})}\BibitemShut {NoStop}%
\bibitem [{\citenamefont {Chambers}(1960)}]{chambers60}%
  \BibitemOpen
  \bibfield  {author} {\bibinfo {author} {\bibfnamefont {R.~G.}\ \bibnamefont
  {Chambers}},\ }\bibfield  {title} {\enquote {\bibinfo {title} {Shift of an
  electron interference pattern by enclosed magnetic flux},}\ }\href {\doibase
  10.1103/PhysRevLett.5.3} {\bibfield  {journal} {\bibinfo  {journal} {Phys.
  Rev. Lett.}\ }\textbf {\bibinfo {volume} {5}},\ \bibinfo {pages} {3}
  (\bibinfo {year} {1960})}\BibitemShut {NoStop}%
\bibitem [{\citenamefont {Matteucci}\ and\ \citenamefont
  {Pozzi}(1985)}]{matteucci85}%
  \BibitemOpen
  \bibfield  {author} {\bibinfo {author} {\bibfnamefont {G.}~\bibnamefont
  {Matteucci}}\ and\ \bibinfo {author} {\bibfnamefont {G.}~\bibnamefont
  {Pozzi}},\ }\bibfield  {title} {\enquote {\bibinfo {title} {New diffraction
  experiment on the electrostatic {A}haronov-{B}ohm effect},}\ }\href {\doibase
  10.1103/PhysRevLett.54.2469} {\bibfield  {journal} {\bibinfo  {journal}
  {Phys. Rev. Lett.}\ }\textbf {\bibinfo {volume} {54}},\ \bibinfo {pages}
  {2469} (\bibinfo {year} {1985})}\BibitemShut {NoStop}%
\bibitem [{\citenamefont {Webb}\ \emph {et~al.}(1985)\citenamefont {Webb},
  \citenamefont {Washburn}, \citenamefont {Umbach},\ and\ \citenamefont
  {Laibowitz}}]{webb85}%
  \BibitemOpen
  \bibfield  {author} {\bibinfo {author} {\bibfnamefont {R.~A.}\ \bibnamefont
  {Webb}}, \bibinfo {author} {\bibfnamefont {S.}~\bibnamefont {Washburn}},
  \bibinfo {author} {\bibfnamefont {C.~P.}\ \bibnamefont {Umbach}}, \ and\
  \bibinfo {author} {\bibfnamefont {R.~B.}\ \bibnamefont {Laibowitz}},\
  }\bibfield  {title} {\enquote {\bibinfo {title} {Observation of $\frac{h}{e}$
  {A}haronov-{B}ohm oscillations in normal-metal rings},}\ }\href {\doibase
  10.1103/PhysRevLett.54.2696} {\bibfield  {journal} {\bibinfo  {journal}
  {Phys. Rev. Lett.}\ }\textbf {\bibinfo {volume} {54}},\ \bibinfo {pages}
  {2696} (\bibinfo {year} {1985})}\BibitemShut {NoStop}%
\bibitem [{\citenamefont {Tonomura}\ \emph {et~al.}(1986)\citenamefont
  {Tonomura}, \citenamefont {Osakabe}, \citenamefont {Matsuda}, \citenamefont
  {Kawasaki}, \citenamefont {Endo}, \citenamefont {Yano},\ and\ \citenamefont
  {Yamada}}]{tonomura86}%
  \BibitemOpen
  \bibfield  {author} {\bibinfo {author} {\bibfnamefont {A.}~\bibnamefont
  {Tonomura}}, \bibinfo {author} {\bibfnamefont {N.}~\bibnamefont {Osakabe}},
  \bibinfo {author} {\bibfnamefont {T.}~\bibnamefont {Matsuda}}, \bibinfo
  {author} {\bibfnamefont {T.}~\bibnamefont {Kawasaki}}, \bibinfo {author}
  {\bibfnamefont {J.}~\bibnamefont {Endo}}, \bibinfo {author} {\bibfnamefont
  {S.}~\bibnamefont {Yano}}, \ and\ \bibinfo {author} {\bibfnamefont
  {H.}~\bibnamefont {Yamada}},\ }\bibfield  {title} {\enquote {\bibinfo {title}
  {Evidence for {A}haronov-{B}ohm effect with magnetic field completely
  shielded from electron wave},}\ }\href {\doibase 10.1103/PhysRevLett.56.792}
  {\bibfield  {journal} {\bibinfo  {journal} {Phys. Rev. Lett.}\ }\textbf
  {\bibinfo {volume} {56}},\ \bibinfo {pages} {792} (\bibinfo {year}
  {1986})}\BibitemShut {NoStop}%
\bibitem [{\citenamefont {van Oudenaarden}\ \emph {et~al.}(1998)\citenamefont
  {van Oudenaarden}, \citenamefont {Devoret}, \citenamefont {Nazarov},\ and\
  \citenamefont {Mooij}}]{oudenaarden98}%
  \BibitemOpen
  \bibfield  {author} {\bibinfo {author} {\bibfnamefont {A.}~\bibnamefont {van
  Oudenaarden}}, \bibinfo {author} {\bibfnamefont {M.~H.}\ \bibnamefont
  {Devoret}}, \bibinfo {author} {\bibfnamefont {Y.~V.}\ \bibnamefont
  {Nazarov}}, \ and\ \bibinfo {author} {\bibfnamefont {J.~E.}\ \bibnamefont
  {Mooij}},\ }\bibfield  {title} {\enquote {\bibinfo {title} {Magneto-electric
  {A}haronov–{B}ohm effect in metal rings},}\ }\href {\doibase 10.1038/35808}
  {\bibfield  {journal} {\bibinfo  {journal} {Nature}\ }\textbf {\bibinfo
  {volume} {391}},\ \bibinfo {pages} {768} (\bibinfo {year}
  {1998})}\BibitemShut {NoStop}%
\bibitem [{\citenamefont {Bachtold}\ \emph {et~al.}(1999)\citenamefont
  {Bachtold}, \citenamefont {Strunk}, \citenamefont {Salvetat}, \citenamefont
  {Bonard}, \citenamefont {Forró}, \citenamefont {Nussbaumer},\ and\
  \citenamefont {Schönenberger}}]{bachtold99}%
  \BibitemOpen
  \bibfield  {author} {\bibinfo {author} {\bibfnamefont {A.}~\bibnamefont
  {Bachtold}}, \bibinfo {author} {\bibfnamefont {C.}~\bibnamefont {Strunk}},
  \bibinfo {author} {\bibfnamefont {J.-P.}\ \bibnamefont {Salvetat}}, \bibinfo
  {author} {\bibfnamefont {J.-M.}\ \bibnamefont {Bonard}}, \bibinfo {author}
  {\bibfnamefont {L.}~\bibnamefont {Forró}}, \bibinfo {author} {\bibfnamefont
  {T.}~\bibnamefont {Nussbaumer}}, \ and\ \bibinfo {author} {\bibfnamefont
  {C.}~\bibnamefont {Schönenberger}},\ }\bibfield  {title} {\enquote {\bibinfo
  {title} {Aharonov–{B}ohm oscillations in carbon nanotubes},}\ }\href
  {\doibase 10.1038/17755} {\bibfield  {journal} {\bibinfo  {journal} {Nature}\
  }\textbf {\bibinfo {volume} {397}},\ \bibinfo {pages} {673} (\bibinfo {year}
  {1999})}\BibitemShut {NoStop}%
\bibitem [{\citenamefont {Ji}\ \emph {et~al.}(2003)\citenamefont {Ji},
  \citenamefont {Chung}, \citenamefont {Sprinzak}, \citenamefont {Heiblum},
  \citenamefont {Mahalu},\ and\ \citenamefont {Shtrikman}}]{ji03}%
  \BibitemOpen
  \bibfield  {author} {\bibinfo {author} {\bibfnamefont {Y.}~\bibnamefont
  {Ji}}, \bibinfo {author} {\bibfnamefont {Y.}~\bibnamefont {Chung}}, \bibinfo
  {author} {\bibfnamefont {D.}~\bibnamefont {Sprinzak}}, \bibinfo {author}
  {\bibfnamefont {M.}~\bibnamefont {Heiblum}}, \bibinfo {author} {\bibfnamefont
  {D.}~\bibnamefont {Mahalu}}, \ and\ \bibinfo {author} {\bibfnamefont
  {H.}~\bibnamefont {Shtrikman}},\ }\bibfield  {title} {\enquote {\bibinfo
  {title} {An electronic {M}ach–{Z}ehnder interferometer},}\ }\href {\doibase
  10.1038/nature01503} {\bibfield  {journal} {\bibinfo  {journal} {Nature}\
  }\textbf {\bibinfo {volume} {422}},\ \bibinfo {pages} {415} (\bibinfo {year}
  {2003})}\BibitemShut {NoStop}%
\bibitem [{\citenamefont {Peng}\ \emph {et~al.}(2010)\citenamefont {Peng},
  \citenamefont {Lai}, \citenamefont {Kong}, \citenamefont {Meister},
  \citenamefont {Chen}, \citenamefont {Qi}, \citenamefont {Zhang},
  \citenamefont {Shen},\ and\ \citenamefont {Cui}}]{peng10}%
  \BibitemOpen
  \bibfield  {author} {\bibinfo {author} {\bibfnamefont {H.}~\bibnamefont
  {Peng}}, \bibinfo {author} {\bibfnamefont {K.}~\bibnamefont {Lai}}, \bibinfo
  {author} {\bibfnamefont {D.}~\bibnamefont {Kong}}, \bibinfo {author}
  {\bibfnamefont {S.}~\bibnamefont {Meister}}, \bibinfo {author} {\bibfnamefont
  {Y.}~\bibnamefont {Chen}}, \bibinfo {author} {\bibfnamefont {X.-L.}\
  \bibnamefont {Qi}}, \bibinfo {author} {\bibfnamefont {S.-C.}\ \bibnamefont
  {Zhang}}, \bibinfo {author} {\bibfnamefont {Z.-X.}\ \bibnamefont {Shen}}, \
  and\ \bibinfo {author} {\bibfnamefont {Y.}~\bibnamefont {Cui}},\ }\bibfield
  {title} {\enquote {\bibinfo {title} {Aharonov–{B}ohm interference in
  topological insulator nanoribbons},}\ }\href {\doibase 10.1038/nmat2609}
  {\bibfield  {journal} {\bibinfo  {journal} {Nat. Mater.}\ }\textbf {\bibinfo
  {volume} {9}},\ \bibinfo {pages} {225} (\bibinfo {year} {2010})}\BibitemShut
  {NoStop}%
\bibitem [{\citenamefont {Becker}\ \emph {et~al.}(2019)\citenamefont {Becker},
  \citenamefont {Guzzinati}, \citenamefont {Béché}, \citenamefont
  {Verbeeck},\ and\ \citenamefont {Batelaan}}]{becker19}%
  \BibitemOpen
  \bibfield  {author} {\bibinfo {author} {\bibfnamefont {M.}~\bibnamefont
  {Becker}}, \bibinfo {author} {\bibfnamefont {G.}~\bibnamefont {Guzzinati}},
  \bibinfo {author} {\bibfnamefont {A.}~\bibnamefont {Béché}}, \bibinfo
  {author} {\bibfnamefont {J.}~\bibnamefont {Verbeeck}}, \ and\ \bibinfo
  {author} {\bibfnamefont {H.}~\bibnamefont {Batelaan}},\ }\bibfield  {title}
  {\enquote {\bibinfo {title} {Asymmetry and non-dispersivity in the
  {A}haronov-{B}ohm effect},}\ }\href {\doibase 10.1038/s41467-019-09609-9}
  {\bibfield  {journal} {\bibinfo  {journal} {Nat. Commun.}\ }\textbf {\bibinfo
  {volume} {10}},\ \bibinfo {pages} {1700} (\bibinfo {year}
  {2019})}\BibitemShut {NoStop}%
\bibitem [{\citenamefont {Nakamura}\ \emph {et~al.}(2019)\citenamefont
  {Nakamura}, \citenamefont {Fallahi}, \citenamefont {Sahasrabudhe},
  \citenamefont {Rahman}, \citenamefont {Liang}, \citenamefont {Gardner},\ and\
  \citenamefont {Manfra}}]{nakamura19}%
  \BibitemOpen
  \bibfield  {author} {\bibinfo {author} {\bibfnamefont {J.}~\bibnamefont
  {Nakamura}}, \bibinfo {author} {\bibfnamefont {S.}~\bibnamefont {Fallahi}},
  \bibinfo {author} {\bibfnamefont {H.}~\bibnamefont {Sahasrabudhe}}, \bibinfo
  {author} {\bibfnamefont {R.}~\bibnamefont {Rahman}}, \bibinfo {author}
  {\bibfnamefont {S.}~\bibnamefont {Liang}}, \bibinfo {author} {\bibfnamefont
  {G.~C.}\ \bibnamefont {Gardner}}, \ and\ \bibinfo {author} {\bibfnamefont
  {M.~J.}\ \bibnamefont {Manfra}},\ }\bibfield  {title} {\enquote {\bibinfo
  {title} {Aharonov–{B}ohm interference of fractional quantum {H}all edge
  modes},}\ }\href {\doibase 10.1038/s41567-019-0441-8} {\bibfield  {journal}
  {\bibinfo  {journal} {Nat. Phys.}\ }\textbf {\bibinfo {volume} {15}},\
  \bibinfo {pages} {563} (\bibinfo {year} {2019})}\BibitemShut {NoStop}%
\bibitem [{\citenamefont {{C. Déprez \it{et al.}}}(2021)}]{deprez21}%
  \BibitemOpen
  \bibfield  {author} {\bibinfo {author} {\bibnamefont {{C. Déprez \it{et
  al.}}}},\ }\bibfield  {title} {\enquote {\bibinfo {title} {A tunable
  {F}abry–{P}érot quantum {H}all interferometer in graphene},}\ }\href
  {\doibase 10.1038/s41565-021-00847-x} {\bibfield  {journal} {\bibinfo
  {journal} {Nature Nanotechnology}\ }\textbf {\bibinfo {volume} {16}},\
  \bibinfo {pages} {555} (\bibinfo {year} {2021})}\BibitemShut {NoStop}%
\bibitem [{\citenamefont {{Y. Ronen \it{et al.}}}(2021)}]{ronen21}%
  \BibitemOpen
  \bibfield  {author} {\bibinfo {author} {\bibnamefont {{Y. Ronen \it{et
  al.}}}},\ }\bibfield  {title} {\enquote {\bibinfo {title} {Aharonov–{B}ohm
  effect in graphene-based {F}abry–{P}érot quantum {H}all
  interferometers},}\ }\href {\doibase 10.1038/s41565-021-00861-z} {\bibfield
  {journal} {\bibinfo  {journal} {Nature Nanotechnology}\ }\textbf {\bibinfo
  {volume} {16}},\ \bibinfo {pages} {563} (\bibinfo {year} {2021})}\BibitemShut
  {NoStop}%
\bibitem [{\citenamefont {Peshkin}\ and\ \citenamefont
  {Lipkin}(1995)}]{peshkin95}%
  \BibitemOpen
  \bibfield  {author} {\bibinfo {author} {\bibfnamefont {M.}~\bibnamefont
  {Peshkin}}\ and\ \bibinfo {author} {\bibfnamefont {H.~J.}\ \bibnamefont
  {Lipkin}},\ }\bibfield  {title} {\enquote {\bibinfo {title} {Topology,
  locality, and {A}haronov-{B}ohm effect with neutrons},}\ }\href {\doibase
  10.1103/PhysRevLett.74.2847} {\bibfield  {journal} {\bibinfo  {journal}
  {Phys. Rev. Lett.}\ }\textbf {\bibinfo {volume} {74}},\ \bibinfo {pages}
  {2847} (\bibinfo {year} {1995})}\BibitemShut {NoStop}%
\bibitem [{\citenamefont {Cohen}\ \emph {et~al.}(2019)\citenamefont {Cohen},
  \citenamefont {Larocque}, \citenamefont {Bouchard}, \citenamefont
  {Nejadsattari}, \citenamefont {Gefen},\ and\ \citenamefont
  {Karimi}}]{cohen19}%
  \BibitemOpen
  \bibfield  {author} {\bibinfo {author} {\bibfnamefont {E.}~\bibnamefont
  {Cohen}}, \bibinfo {author} {\bibfnamefont {H.}~\bibnamefont {Larocque}},
  \bibinfo {author} {\bibfnamefont {F.}~\bibnamefont {Bouchard}}, \bibinfo
  {author} {\bibfnamefont {F.}~\bibnamefont {Nejadsattari}}, \bibinfo {author}
  {\bibfnamefont {Y.}~\bibnamefont {Gefen}}, \ and\ \bibinfo {author}
  {\bibfnamefont {E.}~\bibnamefont {Karimi}},\ }\bibfield  {title} {\enquote
  {\bibinfo {title} {Geometric phase from {A}haronov–{B}ohm to
  {P}ancharatnam–{B}erry and beyond},}\ }\href {\doibase
  10.1038/s42254-019-0071-1} {\bibfield  {journal} {\bibinfo  {journal} {Nat.
  Rev. Phys.}\ }\textbf {\bibinfo {volume} {1}},\ \bibinfo {pages} {437}
  (\bibinfo {year} {2019})}\BibitemShut {NoStop}%
\bibitem [{\citenamefont {Ballentine}(2000)}]{ballentine}%
  \BibitemOpen
  \bibfield  {author} {\bibinfo {author} {\bibfnamefont {L.~E.}\ \bibnamefont
  {Ballentine}},\ }\href@noop {} {\emph {\bibinfo {title} {Quantum mechanics: A
  modern development}}}\ (\bibinfo  {publisher} {World Scientific},\ \bibinfo
  {address} {Singapore},\ \bibinfo {year} {2000})\BibitemShut {NoStop}%
\bibitem [{\citenamefont {{B. H. Bransden}}\ and\ \citenamefont {{C. J.
  Joachain}}(2000)}]{bransden}%
  \BibitemOpen
  \bibfield  {author} {\bibinfo {author} {\bibnamefont {{B. H. Bransden}}}\
  and\ \bibinfo {author} {\bibnamefont {{C. J. Joachain}}},\ }\href@noop {}
  {\emph {\bibinfo {title} {Quantum mechanics}}},\ \bibinfo {edition} {2nd}\
  ed.\ (\bibinfo  {publisher} {Pearson},\ \bibinfo {address} {Harlow},\
  \bibinfo {year} {2000})\BibitemShut {NoStop}%
\bibitem [{\citenamefont {{K. Gottfried}}\ and\ \citenamefont {{T. M.
  Yan}}(2003)}]{gottfried}%
  \BibitemOpen
  \bibfield  {author} {\bibinfo {author} {\bibnamefont {{K. Gottfried}}}\ and\
  \bibinfo {author} {\bibnamefont {{T. M. Yan}}},\ }\href@noop {} {\emph
  {\bibinfo {title} {Quantum mechanics: Fun- damentals}}},\ \bibinfo {edition}
  {2nd}\ ed.\ (\bibinfo  {publisher} {Springer},\ \bibinfo {address} {New
  York},\ \bibinfo {year} {2003})\BibitemShut {NoStop}%
\bibitem [{\citenamefont {Saldanha}(2023)}]{saldanha23}%
  \BibitemOpen
  \bibfield  {author} {\bibinfo {author} {\bibfnamefont {P.~L.}\ \bibnamefont
  {Saldanha}},\ }\bibfield  {title} {\enquote {\bibinfo {title} {Electrodynamic
  {A}haronov-{B}ohm effect},}\ }\href {\doibase 10.1103/PhysRevA.108.062218}
  {\bibfield  {journal} {\bibinfo  {journal} {Phys. Rev. A}\ }\textbf {\bibinfo
  {volume} {108}},\ \bibinfo {pages} {062218} (\bibinfo {year}
  {2023})}\BibitemShut {NoStop}%
\bibitem [{\citenamefont {Singleton}\ and\ \citenamefont
  {Vagenas}(2013)}]{singleton13}%
  \BibitemOpen
  \bibfield  {author} {\bibinfo {author} {\bibfnamefont {D.}~\bibnamefont
  {Singleton}}\ and\ \bibinfo {author} {\bibfnamefont {E.~C.}\ \bibnamefont
  {Vagenas}},\ }\bibfield  {title} {\enquote {\bibinfo {title} {The covariant,
  time-dependent {A}haronov–{B}ohm effect},}\ }\href {\doibase
  https://doi.org/10.1016/j.physletb.2013.05.014} {\bibfield  {journal}
  {\bibinfo  {journal} {Phys. Lett. B}\ }\textbf {\bibinfo {volume} {723}},\
  \bibinfo {pages} {241} (\bibinfo {year} {2013})}\BibitemShut {NoStop}%
\bibitem [{\citenamefont {Santos}\ and\ \citenamefont
  {Gonzalo}(1999)}]{santos99}%
  \BibitemOpen
  \bibfield  {author} {\bibinfo {author} {\bibfnamefont {E.}~\bibnamefont
  {Santos}}\ and\ \bibinfo {author} {\bibfnamefont {I.}~\bibnamefont
  {Gonzalo}},\ }\bibfield  {title} {\enquote {\bibinfo {title} {Microscopic
  theory of the {A}haronov-{B}ohm effect},}\ }\href {\doibase
  10.1209/epl/i1999-00182-9} {\bibfield  {journal} {\bibinfo  {journal} {EPL}\
  }\textbf {\bibinfo {volume} {45}},\ \bibinfo {pages} {418} (\bibinfo {year}
  {1999})}\BibitemShut {NoStop}%
\bibitem [{\citenamefont {Marletto}\ and\ \citenamefont
  {Vedral}(2020)}]{marletto20}%
  \BibitemOpen
  \bibfield  {author} {\bibinfo {author} {\bibfnamefont {C.}~\bibnamefont
  {Marletto}}\ and\ \bibinfo {author} {\bibfnamefont {V.}~\bibnamefont
  {Vedral}},\ }\bibfield  {title} {\enquote {\bibinfo {title} {Aharonov-{B}ohm
  phase is locally generated like all other quantum phases},}\ }\href {\doibase
  10.1103/PhysRevLett.125.040401} {\bibfield  {journal} {\bibinfo  {journal}
  {Phys. Rev. Lett.}\ }\textbf {\bibinfo {volume} {125}},\ \bibinfo {pages}
  {040401} (\bibinfo {year} {2020})}\BibitemShut {NoStop}%
\bibitem [{\citenamefont {Saldanha}(2021{\natexlab{a}})}]{saldanha21a}%
  \BibitemOpen
  \bibfield  {author} {\bibinfo {author} {\bibfnamefont {P.~L.}\ \bibnamefont
  {Saldanha}},\ }\bibfield  {title} {\enquote {\bibinfo {title} {Local
  description of the {A}haronov–{B}ohm effect with a quantum electromagnetic
  field},}\ }\href {\doibase 10.1007/s10701-021-00414-3} {\bibfield  {journal}
  {\bibinfo  {journal} {Found. Phys.}\ }\textbf {\bibinfo {volume} {51}},\
  \bibinfo {pages} {6} (\bibinfo {year} {2021}{\natexlab{a}})}\BibitemShut
  {NoStop}%
\bibitem [{\citenamefont {Saldanha}(2021{\natexlab{b}})}]{saldanha21b}%
  \BibitemOpen
  \bibfield  {author} {\bibinfo {author} {\bibfnamefont {P.~L.}\ \bibnamefont
  {Saldanha}},\ }\bibfield  {title} {\enquote {\bibinfo {title}
  {Aharonov-{C}asher and shielded {A}haronov-{B}ohm effects with a quantum
  electromagnetic field},}\ }\href {\doibase 10.1103/PhysRevA.104.032219}
  {\bibfield  {journal} {\bibinfo  {journal} {Phys. Rev. A}\ }\textbf {\bibinfo
  {volume} {104}},\ \bibinfo {pages} {032219} (\bibinfo {year}
  {2021}{\natexlab{b}})}\BibitemShut {NoStop}%
\bibitem [{\citenamefont {Saldanha}(2024)}]{saldanha24}%
  \BibitemOpen
  \bibfield  {author} {\bibinfo {author} {\bibfnamefont {P.~L.}\ \bibnamefont
  {Saldanha}},\ }\bibfield  {title} {\enquote {\bibinfo {title} {Gauge
  invariance of the {A}haronov-{B}ohm effect in a quantum electrodynamics
  framework},}\ }\href {\doibase 10.1103/PhysRevA.109.062205} {\bibfield
  {journal} {\bibinfo  {journal} {Phys. Rev. A}\ }\textbf {\bibinfo {volume}
  {109}},\ \bibinfo {pages} {062205} (\bibinfo {year} {2024})}\BibitemShut
  {NoStop}%
\bibitem [{\citenamefont {Batelaan}\ and\ \citenamefont
  {Tonomura}(2009)}]{batelaantonomura}%
  \BibitemOpen
  \bibfield  {author} {\bibinfo {author} {\bibfnamefont {H.}~\bibnamefont
  {Batelaan}}\ and\ \bibinfo {author} {\bibfnamefont {A.}~\bibnamefont
  {Tonomura}},\ }\bibfield  {title} {\enquote {\bibinfo {title} {{The
  {A}haronov–{B}ohm effects: Variations on a subtle theme}},}\ }\href
  {\doibase 10.1063/1.3226854} {\bibfield  {journal} {\bibinfo  {journal}
  {Physics Today}\ }\textbf {\bibinfo {volume} {62}},\ \bibinfo {pages} {38}
  (\bibinfo {year} {2009})}\BibitemShut {NoStop}%
\end{thebibliography}
%

\end{document}